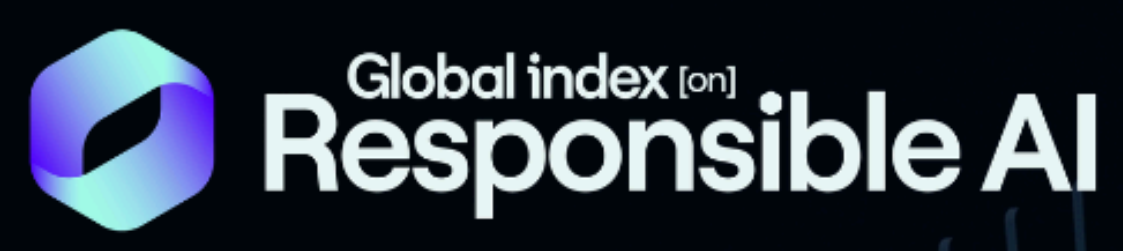

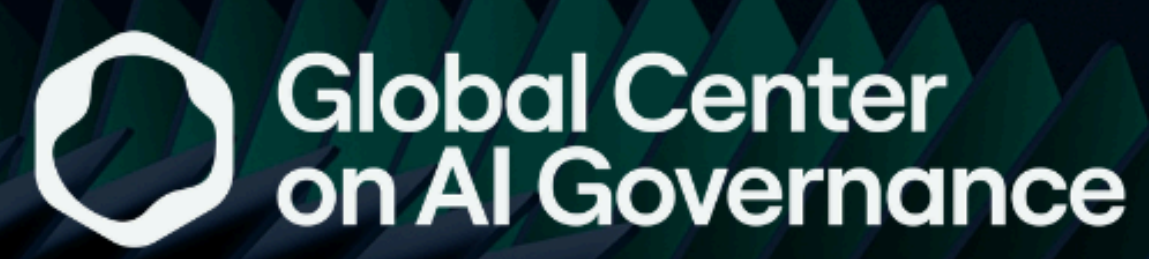

# Global Index on Responsible AI

## Conceptual Framework and Methodology

Fola Adeleke
Rachel Adams
Ayantola Alayande
Daniela Benavente
Ana Florido
Nicolás Grossman
Leah Junck

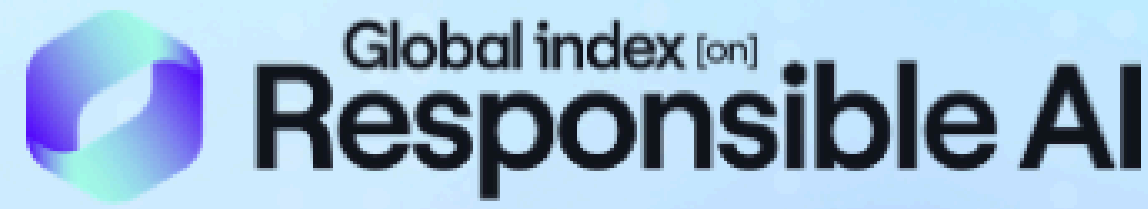

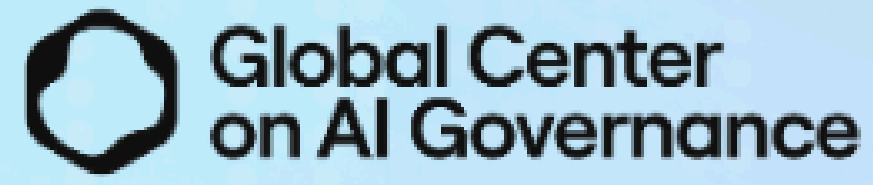

## Table of Contents

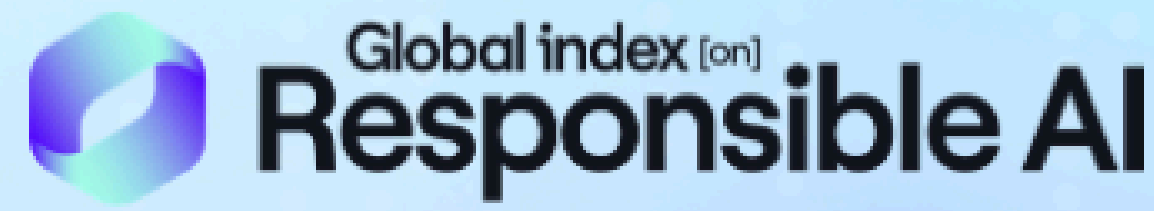
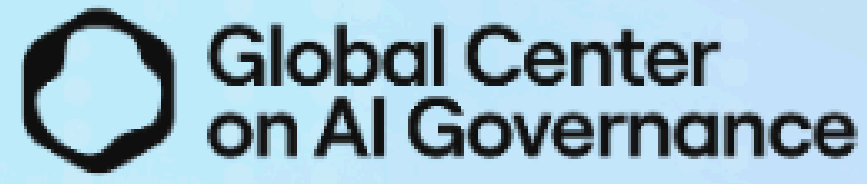

## 1. INTRODUCTION

### 1.1 Why measuring Responsible AI

The widespread adoption of artificial intelligence (AI) across multiple economic sectors[1], and its increasing direct and systemic impacts on individuals and communities across the world, has made clear the need to govern this pervasive technology. Governing AI means ensuring that it benefits society and can be applied to help meet the Sustainable Development Goals[2], while simultaneously preventing and mitigating the harms it might cause. This need has been recognized in multiple global fora, international treaties, and regional and national initiatives.

Over the past decade, a wide range of international, regional, and national initiatives have sought to establish principles, norms, and governance approaches for AI. Although these initiatives differ in scope and emphasis, they share a common objective: ensuring that the design, development, deployment, adoption, and governance of AI systems are aligned with human rights, democratic values, and ethical principles. This objective is reflected in a growing body of United Nations agreements and policy frameworks—such as the UNESCO *Recommendation on the Ethics of Artificial Intelligence (UNESCO Recommendation)*[3], and the *Global Digital Compact*[4]—which outline guiding principles for international cooperation to harness the benefits of digital technologies to achieve Sustainable Development Goals and advance all human rights while reducing existing inequalities between and within countries. It is further strengthened by *Governing AI for Humanity*[5], a UN AI Advisory Body report that strengthens the case for global governance approaches to complement local and regional initiatives, and identifies key gaps to prioritize action.

It is important to note that all these global agreements and approaches emanate from the principles of the Charter of the United Nations, international human rights law[6], the 2030 Agenda for Sustainable Development[7], and the UN Guiding Principles on Business and Human Rights[8], and can be understood as ways to bring these into the realm of new technologies, specifically AI[9]. Therefore, proper and effective governance of AI technologies is key to "advance all human rights, including the rights of the child, the rights of persons with disabilities, and the right to development," as stated in the Global Digital Compact.

[1] See Filippucci, F. et al. (2024), "The impact of Artificial Intelligence on productivity, distribution and growth: Key mechanisms, initial evidence and policy challenges", OECD Artificial Intelligence Papers, No. 15, OECD Publishing, Paris, https://doi.org/10.1787/8d900037-en.
[2] https://sdgs.un.org/goals
[3] United Nations Educational, Scientific and Cultural Organization, Recommendation on the Ethics of Artificial Intelligence (Paris: UNESCO, 2021), accessed July 1, 2026, https://unesdoc.unesco.org/ark:/48223/pf0000380455.
[4] United Nations. (2024). Global Digital Compact. United Nations. https://www.un.org/global-digital-compact/en
[5] United Nations Secretary-General's High-Level Advisory Body on Artificial Intelligence. (2023). Governing AI for humanity: Interim report. United Nations. https://www.un.org/sites/un2.un.org/files/un_ai_advisory_body_governing_ai_for_humanity_interim_report.pdf
[6] United Nations General Assembly. (1948). Universal Declaration of Human Rights (General Assembly resolution 217 A (III)). United Nations. https://www.un.org/en/about-us/universal-declaration-of-human-rights
[7] United Nations General Assembly. (2015). Transforming our world: The 2030 Agenda for Sustainable Development (A/RES/70/1). United Nations. https://sdgs.un.org/2030agenda
[8] United Nations Human Rights Council. (2011). Guiding Principles on Business and Human Rights: Implementing the United Nations "Protect, Respect and Remedy" framework. Office of the High Commissioner for Human Rights. https://www.ohchr.org/sites/default/files/documents/publications/guidingprinciplesbusinesshr_en.pdf

Yet, commitments alone are insufficient. Responsible AI cannot be secured through principles and aspirations alone. It requires institutions, governance mechanisms, public participation, oversight arrangements, implementation capacity, monitoring systems, and accessible pathways for accountability and redress. Without mechanisms to assess whether these elements exist and function in practice, responsible AI remains difficult to evaluate and compare across countries.

Measurement therefore plays a critical role in advancing responsible AI. By systematically collecting evidence on governance frameworks, implementation efforts, civil society engagement, and enabling conditions, measurement makes it possible to assess how countries are translating commitments into practice, and to identify areas where further action is needed.

### 1.2 GIRAI in the Current Landscape of AI Composite Indicators

The Global Index on Responsible AI (GIRAI) was established to address the lack of globally comparable evidence on responsible AI governance. Prior to the development of the Index, there was limited systematic information on how countries were responding to the opportunities and risks created by AI, particularly from a human rights perspective. Existing initiatives provided valuable assessments of national readiness and institutional capacity, but there remained a need for a global measurement framework capable of systematically documenting and comparing responsible AI governance across countries.

Several initiatives already assess different aspects of countries' AI development, readiness, capacity, and governance. These include the Stanford AI Index[10], which tracks global trends in AI research, development, adoption, investment, and impact; the Tortoise Global AI Index[11], which measures national AI capacity through indicators related to implementation, innovation, and investment; the Oxford Insights Government AI Readiness Index[12], which evaluates governments' readiness to use AI in the public sector; and the IMF AI Preparedness Index[13], which focuses on countries' preparedness for AI adoption across areas such as digital infrastructure, human capital, innovation, economic integration, and regulation.

[9] The advantages of rooting AI Governance frameworks through a Human Rights lens has also been explored and highlighted by several scholars. Donahoe, E., & Metzger, M. M. (2019) for example, advocate for this perspective, not only for the practicality of building on top of already globally accepted values and principles, which in many cases are also recognized in national legislations, but also because the human rights framework, as developed in the Universal Declaration of Human Rights, specifically addresses key ethical concerns around AI and challenges such as bias and nondiscrimination (Art. 2), the right to life and security (Art. 3), due process and remedies (Arts. 8–11), privacy (Art. 12), freedom of expression (Art. 19), democratic participation (Arts. 20–21), and the right to work and adequate living standards (Arts. 23, 25).

[10] Stanford HAI (2025). AI Index Report 2026. Stanford, CA: Stanford University. https://hai.stanford.edu/assets/files/ai_index_report_2026.pdf
[11] Tortoise Media (2024). Global AI Index. London: Tortoise Media. https://www.tortoisemedia.com/data/global-ai
[12] Oxford Insights (2025). Government AI Readiness Index. Oxford: Oxford Insights, https://oxfordinsights.com/wp-content/uploads/2026/06/LATEST-GAIRI-Report-2025.pdf

Other initiatives focus on AI governance and values. The Artificial Intelligence and Democratic Values Index examines the extent to which national AI policies and practices align with democratic values and international governance frameworks. The OECD.AI Index[14] monitors progress towards the implementation of the OECD Recommendation on Artificial Intelligence through indicators covering policy environments, enabling infrastructure, research and development, jobs and skills, and international cooperation.

GIRAI complements these initiatives which address AI capacity, innovation performance, readiness, or ecosystem development, by focusing specifically on responsible AI governance from a human rights-based perspective. It assesses whether countries are establishing governance frameworks, implementation measures, civil society engagement mechanisms, and enabling conditions associated with Responsible AI. By combining primary and secondary data, the index provides a structured and globally comparable assessment of how countries are translating Responsible AI principles into governance commitments and practices.

### 1.3 Scope and Purpose of the 2026 Edition

The 2026 edition represents the second edition of GIRAI. Building on the experience and lessons learned from the first edition, the measurement framework has been refined to improve conceptual clarity, strengthen the assessment of implementation, consider emergent risks, and enhance the consistency and comparability of the data collected.

The index combines evidence on AI policy frameworks and initiatives, civil society engagement, and broader enabling conditions to provide a comprehensive assessment of responsible AI governance. By generating comparable evidence across countries and regions, GIRAI supports policymakers, researchers, civil society organisations (CSOs), international institutions, and other stakeholders seeking to understand the current state of responsible AI governance and identify opportunities for improvement.

The GIRAI conceptual framework consists of 38 indicators organised into five dimensions and three pillars, in a matrix format. Countries are assessed across five dimensions of responsible AI: Inclusion and Diversity; Ethics and Sustainability; Labour and Skills; Trust and Safety; and AI Use in Public Service. Within each dimension, performance is measured through three complementary pillars: AI Policy, CSO Engagement, and Enabling Conditions. In addition, the Index separately assesses evidence of government deployment of Unacceptable Risk AI Systems (URAI), which is incorporated as an accountability penalty applied to the final score.

[13] IMF (2024). AI Preparedness Index. Washington, DC: International Monetary Fund. https://www.imf.org/external/datamapper/datasets/AIPI
[14] OECD (2026), The OECD.AI Index, OECD Publishing, Paris, https://doi.org/10.1787/32c01014-en.

*Table 1: GIRAI's Conceptual Framework (Dimensions, Pillars, and Indicators)*

| GIRAI Matrix<br><br>Dimensions (rows), Pillars (columns), and indicators (cells) | **A: AI Policy**<br><br>For each indicator,<br><br>AI Policy = Frameworks + Initiatives | **B: CSO Engagement** | **C: Enabling Conditions** |
|---|---|---|---|
| 1 Inclusion and Diversity | 1. Gender Equality<br><br>2. Children's Rights<br><br>3. Cultural and Linguistic Diversity | 1. Civil Society Engagement in Inclusion and Diversity | *1. Egalitarian Democracy*<br><br>*2. Device Affordability*<br><br>*3. Gender Gap in Mobile Internet* |
| 2 Ethics and Sustainability | 4. Fairness and Non-Discrimination<br><br>5. Transparency and Explainability<br><br>6. Human Oversight and Determination<br><br>7. Environmental Impact | 2. Civil Society Engagement in Ethics and Sustainability | *4. Rule of Law*<br><br>*5. Low-Carbon Energy Share* |
| 3 Labour and Skills | 8. Labour Protections<br><br>9. Reskilling and Upskilling Initiatives<br><br>10. AI Literacy | 3. Civil Society Engagement in Labour and Skills | *6. Skills and Literacy*<br><br>*7. Labour Rights Compliance* |
| 4 Trust and Safety | 11. Safety and Security<br><br>12. Access to Redress and Remedy<br><br>13. Impact Assessments<br><br>14. AI-facilitated Misinformation and Violence | 4. Civil Society Engagement in Trust and Safety | *8. Cybersecurity*<br><br>*9. Data Protection and Privacy*<br><br>*10. Data Sharing and Access*<br><br>*11. Consumer Protection*<br><br>*12. Global Peace* |
| 5 AI Use in Public Service | 15. Public Sector Skills Development<br><br>16. Public Disclosure of Government | 5. Government Mechanisms for CSO Inclusion in AI Policy | *13. Civil Society Accountability* |

| | Algorithmic Systems<br><br>17. Public Procurement | and Governance | *14. Public Service Delivery*<br><br>*15. Access to Public Information* |
|---|---|---|---|
| Unacceptable Risks AI Systems (URAI penalty) - Transversal | | | |

The purpose of GIRAI is not only to rank countries, but also to provide a structured evidence base for understanding the strengths, weaknesses, and development trajectories of responsible AI governance across countries[15]. By documenting both policy commitments and evidence of implementation, the index seeks to support informed decision-making, encourage accountability, and contribute to ongoing efforts to promote responsible AI governance globally.

The framework was informed by a systematised review of AI governance frameworks and by consultations with regional partner organisations, researchers, and experts participating in regional and global AI governance fora. The second edition further incorporates lessons learned from the first edition of the index, including refinements to the conceptual framework, data collection instruments, indicator guidance, and scoring methodology. Proposed revisions were discussed through consultations with members of the Global Research Network and reviewed by the GIRAI Scientific Advisory Committee as part of the methodological development process.

## 2. Conceptual Framework

### 2.1 Responsible AI

GIRAI was developed as a human rights-based measurement initiative. The index operationalises values, principles, and policy areas derived from the UNESCO Recommendation and related international frameworks in order to assess their implementation at the country level. Through the collection of primary and secondary data, GIRAI seeks to measure the extent to which countries are establishing the governance conditions necessary for responsible AI.

Responsible AI refers to the design, development, deployment, adoption, and governance of artificial intelligence in ways that respect human rights, protect people from harm, promote inclusion, and serve the public good. It reflects the recognition that AI systems are not neutral technologies, but socio-technical systems that are developed, deployed, and governed within existing social, economic, political, and cultural contexts.

[15] Throughout this report, we use the term "countries" or "jurisdictions" interchangeably, to simplify communication. We recognise that some of the jurisdictions included in the GIRAI are currently disputed in nature. The naming of these jurisdictions in this report, as well as their geographical representation on the GIRAI website, does not reflect any position on such disputes or the views of the organisation.

As AI systems increasingly shape access to information, public services, employment opportunities, education, healthcare, and participation in public life, ensuring that these technologies are aligned with human rights and democratic values has become a central governance challenge. Responsible AI therefore requires not only technical safeguards, but also institutions, policies, oversight mechanisms, accountability arrangements, and meaningful public participation.

The overarching assumption of the GIRAI is that design, development, deployment, adoption, and governance of AI systems by all actors must align with established human rights frameworks and processes to harness the benefits of these technologies, while protecting people from potential risks, whether they are newly created by these technologies, or an exacerbation of existing ones. However, governments have a dual role: they are both regulators—responsible for drafting, approving, and implementing policies, laws, and other instruments—and adopters of AI systems. The use of AI by governments requires particular scrutiny, given their obligations for transparency and the privileged access they hold to sensitive data. CSOs play a critical role in promoting and protecting human rights in the context of AI, due to their active involvement in policy creation and their oversight of the deployment of these technologies both by private and government actors.

*Table 2* presents the alignment between the GIRAI framework and the UNESCO Recommendation.

*Table 2. Mapping of GIRAI Indicators to the UNESCO Recommendation on the Ethics of Artificial Intelligence*

| **Principles - UNESCO Recommendation** | **GIRAI Indicator(s)** |
|---|---|
| Proportionality & Do No Harm – AI must only be used for legitimate aims, with thorough risk assessment. | Impact assessments, Government Unacceptable Risk use of AI, Environmental Impact |
| Safety & Security – Prevent technical failures and vulnerabilities throughout the AI lifecycle. | Safety and Security |
| Right to Privacy & Data Protection – Privacy must be ensured at every stage. | Data protection and privacy (secondary data) |
| Multi-stakeholder & Adaptive Governance – Include diverse, evolving participation in governance. | Government Mechanisms for CSO Inclusion in AI Policy and Governance |
| Responsibility & Accountability – Systems should be auditable, traceable, and subject to oversight. | Transparency and Explainability, Access to Redress and Remedy, Public Disclosure of Government Algorithmic Systems |
| Transparency & Explainability – Ensure context-appropriate clarity about how AI works. | Transparency and Explainability |
| Human Oversight & Determination – Maintain human decision-making, not just algorithmic. | Human Oversight and Determination |
| Sustainability – AI systems should align with evolving goals like the SDGs. | Environmental Impact, Gender Equality, Children's rights, Cultural and Linguistic Diversity, AI Literacy, |
| Awareness & Literacy – Promote public AI knowledge through education and civic engagement. | AI Literacy |
| Fairness & Non-Discrimination – Ensure AI is inclusive, just, and combats inequality. | Fairness and Non-Discrimination |
| | |
| **Policy areas - UNESCO Recommendation** | |
| Ethical Impact Assessment – Implement processes to evaluate AI's societal risks/benefits. | Impact Assessments |
| Ethical Governance & Stewardship – Build inclusive, transparent governance frameworks. | Access to Public Information (secondary data) |
| Data Governance – Ensure quality, sovereignty, privacy, and responsible use of data. | Data protection and privacy, Data sharing and access (secondary data) |
| Development & International Cooperation – Foster equitable access and global partnerships. | - |
| Environment & Ecosystems – Track and reduce AI's ecological footprint. | Environmental Impact |
| Gender – Promote gender equality and address AI-related gender biases. | Gender Equality |
| Culture – Respect and embed cultural and linguistic diversity in AI systems. | Cultural and Linguistic Diversity |
| Education & Research – Expand AI literacy, education, and research capabilities. | AI Literacy |
| Communication & Information – Support freedom of expression and information access. | AI-facilitated Misinformation and Violence, Access to Public Information (Secondary data) |
| Economy & Labor – Tackle AI's effects on employment and economic inclusivity. | Labour protections, Reskilling and upskilling initiatives |
| Health & Social Well-Being – Use AI to enhance health outcomes and societal welfare. | - |

**2.2 Index structure**

The GIRAI conceptual framework is organised around five dimensions, three pillars, and 38 indicators. Together, these components provide a structured approach for assessing countries' commitments and practices related to Responsible AI.

**2.3 GIRAI Dimensions**

GIRAI is organised around five dimensions that capture distinct areas of Responsible AI : Inclusion and Diversity, Ethics and Sustainability, Labour and Skills, Trust and Safety, and AI Use in Public Service.

Dimensions measure whether countries are building the frameworks, institutions, practices, and enabling conditions needed to govern AI responsibly and serve as broad organising principles for responsible AI:

- **Inclusion and Diversity**: AI must be governed so that it does not reproduce existing hierarchies, deepen exclusion, or concentrate benefits among those already advantaged. This dimension assesses whether countries are protecting groups and communities most likely to be marginalised or harmed by AI systems.
- **Ethics and Sustainability**: AI must be developed and deployed in ways that are fair, accountable, rights-respecting, and environmentally sustainable. This dimension assesses whether countries are addressing the ethical and ecological conditions under which AI is built and used.
- **Labour and Skills**: AI is reshaping work, skills, livelihoods, and labour rights. This dimension assesses whether countries are preparing people to participate in AI-enabled economies while also protecting workers from displacement, exploitation, surveillance, discrimination, and weakened collective rights.
- **Trust and Safety**: AI systems must be safe, secure, reliable, and accountable if they are to serve the public good. This dimension assesses whether countries are putting in place safeguards to prevent harm, respond to misuse, and maintain public trust.
- **AI Use in Public Service**: When governments use AI, they exercise public power through technical systems. This dimension assesses whether countries are governing public-sector AI in ways that uphold rights, democratic values, transparency, accountability, and access to essential services.

**2.4 GIRAI Pillars**

The GIRAI framework assesses each dimension through three complementary pillars: AI Policy, CSO Engagement, and Enabling Conditions.

These pillars capture different but interrelated components of responsible AI governance. Together, they provide a multidimensional assessment of countries' commitments and practices related to Responsible AI. The first two pillars are entirely based on primary data, whereas the third one uses secondary data.

- The **AI Policy** pillar assesses government action to advance responsible AI. It captures evidence of policy frameworks and government-led initiatives addressing issues related to each of the five dimensions and examines whether governments have established frameworks and undertaken actions to address specific responsible AI issues, and demonstrated evidence of implementation.
- The **CSO Engagement** pillar assesses the contribution of civil society organisations to responsible AI. Civil society organisations play an important role in promoting public participation, supporting accountability, advocating for rights, identifying emerging risks, and representing the interests of affected communities.
- The **Enabling Conditions** pillar assesses broader country-level conditions associated with responsible AI governance. These conditions may support or constrain the ability of governments, civil society organisations, and other actors to advance responsible AI. The pillar captures conditions related to democratic governance, rights protection, access to information, digital inclusion, labour rights, environmental sustainability, cybersecurity, and public sector capacity, among others.

Together, these pillars provide a multidimensional assessment of countries' commitments and practices related to responsible AI. From a policy perspective, each pillar refers to different agents and stakeholders, government officials for AI Policy, CSOs engaged in AI advocacy for CSO Engagement, and the country at large for Enabling Conditions.

## 3. Changes in the Second Edition

### 3.1 Overview

The 2026 edition builds on the conceptual and methodological foundations established in the first edition of the GIRAI. While the overall objective of the index remains unchanged, the second edition introduces refinements to both the conceptual framework and the measurement methodology. These revisions aim to enhance conceptual clarity, strengthen the assessment of implementation, and provide a more comprehensive and nuanced appraisal of responsible AI governance while improving the ability of the index to capture differences in the depth, scope, and implementation of governance responses.

### 3.2 Revisions to the Conceptual Framework

While the first edition was organised around three dimensions, nineteen thematic areas, and three pillars, the second edition introduces a revised structure consisting of five dimensions, three pillars, and thirty-eight indicators.

The revised framework also adopts a new matrix structure. In the first edition, pillars were calculated within each thematic area. In the second edition, each indicator belongs to a specific pillar and aggregation begins at the pillar/dimension level.

The first two pillars of the first edition, Government Frameworks and Government Actions, were merged into the AI Policy pillar, which now assesses both government frameworks and government-led initiatives. The third pillar of the first edition, Non-State Actors, was narrowed to CSO Engagement. Ineffect, while the first edition assessed activities undertaken by a range of non-state actors, including civil society organisations, academia, and the private sector, the second edition focuses specifically on CSOs. This revision reflects the central role of civil society in promoting accountability, participation, and rights-based approaches to AI governance, while also reducing the data collection burden on in-country researchers.

The second edition also changes how enabling conditions are incorporated into the framework. In the first edition, contextual indicators were used as coefficients to adjust pillar scores. In the second edition, these indicators are embedded directly within the conceptual framework, in the third pillar, Enabling Conditions. This approach improves interpretability and recognises that responsible AI governance is shaped not only by policy interventions, but also by broader institutional, social, economic, and governance conditions.

Rather than assessing each thematic area through identical pillar structures, the second edition organises indicators across dimensions and pillars according to the specific responsible AI governance issue they address.

### 3.3 Revisions to the Measurement Framework

The second edition strengthens the assessment of governance frameworks by moving beyond an existence-based approach towards a more detailed assessment of their characteristics and implementation potential.

In the first edition, framework assessment focused primarily on the existence of a framework and its level of enforceability derived from its type. The revised methodology expands the assessment to capture additional characteristics of governance instruments and provides a more granular picture of policy responses.

In addition to framework status (adopted vs draft) and enforceability, the second edition incorporates the assessment of framework reach (who it applies to), sectoral scope (horizontal / vertical), defence and security exemptions, and thematic coverage. The revised methodology also incorporates a dedicated assessment of consultation processes associated with framework development.

In addition, the second edition assesses operationalisation by considering whether frameworks include provisions that support implementation, such as implementation bodies or mechanisms, implementation plans, budgetary estimates, and monitoring and evaluation arrangements.

Together, these revisions provide a more detailed assessment of the content, scope, and implementation potential of responsible AI frameworks.

### 3.4 Strengthening the Assessment of Implementation

A central objective of the second edition is to strengthen the assessment of implementation.

The first edition included Government Actions as a dedicated pillar to capture evidence of implementation. The second edition builds on this approach by integrating implementation more closely into the assessment of AI Policy and expanding the range of implementation evidence captured by the index.

To support this objective, the revised methodology increases the number of government-led initiatives that may be assessed for each indicator, from one Government Action in the first edition to up to three initiatives in the second edition. This allows the index to capture a broader range of implementation efforts and provides a more detailed picture of government action across responsible AI issues.

The second edition also revises the treatment of draft frameworks. In the first edition, draft instruments were assessed as Government Actions. In the revised methodology, publicly available draft frameworks may be assessed as frameworks when there is evidence that they remained active during the study period, such as public consultation processes, parliamentary discussion, official review, or other documented forms of policy development activity. This change recognises the role that policy development processes can play in shaping responsible AI governance prior to formal adoption.

### 3.5 Comparability Across Editions

The methodological revisions introduced in the second edition limit the direct comparability of GIRAI scores across editions, meaning that differences in scores may reflect both changes in country performance and changes in the measurement framework itself. For this reason, comparisons between editions do not rely on overall GIRAI scores. Instead, comparisons on indicators and types of evidence can be meaningfully mapped across both editions.

Comparisons are restricted to equivalent forms of evidence. For frameworks, comparable measures include framework existence, framework type, and framework enforceability. For government-led initiatives and CSO initiatives, comparisons are limited to the existence of at least one initiative associated with the indicator under review. Indicators from the first edition were reviewed and mapped to the corresponding indicators in the second edition, allowing comparable evidence to be analysed within a common framework. Where indicators were substantially revised between editions, evidence from the first edition was reassessed and classified according to the current indicator definitions.

Table 3 presents the mapping between indicators from the first and second editions used for comparative analysis.

*Table 3: Comparison framework between 2024 and 2026 GIRAI editions.*

| | **2026 indicator** | **2024 indicator** |
|---|---|---|
| 1 | Access to Redress and Remedy | Access to Remedy and Redress |
| 2 | Fairness and Non-discrimination | Bias and Unfair Discrimination |
| 3 | Children’s Rights | Children’s Rights |
| 4 | Cultural and Linguistic Diversity | Cultural and Linguistic Diversity |
| 5 | Gender Equality | Gender Equality |
| 6 | Human Oversight and Determination | Human Oversight and Determination |
| 7 | Impact Assessments | Impact Assessments |
| 8 | Labour Protections | Labour Protection and Right to Work |
| 9 | Reskilling and upskilling initiatives | Labour Protection and Right to Work |

| | 2026 indicator | 2024 indicator |
|---|---|---|
| 10 | Public Procurement | Public Procurement |
| 11 | Public Sector Skills Development | Public Sector Skills Development |
| 12 | Safety and Security | Safety, Accuracy and Reliability |
| 13 | Transparency and Explainability | Transparency and Explainability |
| 14 | Government Mechanisms for CSO Inclusion in AI Policy and Governance | National AI Policy |

Two important methodological adjustments were required. First, Government Mechanisms for CSO Inclusion in AI Policy and Governance was mapped to National AI Policy from the first edition, as the relevant framework assessed under the current indicator is the national AI policy or equivalent AI governance framework. Second, Labour Protection and Right to Work was mapped to both Labour Protections and Reskilling and Upskilling Initiatives in the second edition. To enable cross-edition comparison, evidence collected under that indicator was re-assessed and re-classified according to the 2026 indicator definitions: evidence could be assigned to Labour Protections, to Reskilling and Upskilling Initiatives, or to both where it addressed both indicators.

Using this approach, it is possible to analyse changes in framework coverage, framework enforceability, and the existence of government-led and CSO initiatives across editions while maintaining consistency with the revised conceptual framework.

A complete tool for exploring edition-to-edition comparison is available at the GIRAI webpage.

### 4. Data Collection and Sources

#### 4.1 Overview

The GIRAI combines primary and secondary data sources. Primary data are collected through the Global Survey and provide evidence for the AI Policy and CSO Engagement pillars, as well as the assessment of government deployment of Unacceptable Risk AI Systems (URAI). Secondary data sources are used to construct the Enabling Conditions pillar.

#### 4.2 Country Coverage and Study Timeframe

The 2026 edition of the GIRAI covers 135 countries and jurisdictions *(Figure 1).* The countries and jurisdictions included in the second edition were selected based on the availability of suitable country researchers and the existing network of institutional partners supporting the research process. Governments were not involved in determining whether their country was included in the index. Primary data collection covered the period from 1 November 2023 to 30 September 2025.

*Figure 1. GIRAI Country Coverage*

Yes No

Source: World Bank Official Boundaries

#### 4.3 Research Network

The 2026 edition of GIRAI relied on a global network of more than 160 in-country researchers, reviewers, coordinators, supervisors, and institutional partners.

Country researchers were responsible for identifying and assessing evidence relevant to the indicators covered by the Global Survey *(Section 5)*. The use of in-country researchers enabled the index to draw on local knowledge of legal systems, governance arrangements, policy processes, and civil society landscapes. At the same time, standardised coding guidance, validation procedures, and review mechanisms supported consistency across countries.

All survey instruments, methodological guidance, validation procedures, and reviewer communications were conducted in English. However, evidence collection was carried out using original-language sources. Researchers identified and assessed documentation in relevant national languages and recorded evidence according to standardised coding procedures.

This approach was designed to ensure that country assessments were grounded in local expertise while adhering to common methodological standards across all participating countries.

#### 4.4 Validation and Quality Assurance

The GIRAI methodology relies on a multi-layer quality assurance process designed to promote consistency, transparency, and reliability across countries.

Country researchers identify evidence, complete surveys, document sources, and provide justification for coding decisions. Evidence was web-archived and in the case of documents, retained through a central repository to support transparency, verification, and future review.

Country Coordinators or Regional Supervisors conduct a first layer of review, assessing the completeness of submissions, the supporting evidence, and identifying potential omissions. Corrections are requested where necessary. Following first-level review, submissions undergo independent assessment by Global Reviewers, who assess both the coding decisions and the supporting evidence provided by researchers. Questionnaires may be returned for revision multiple times before approval. Finally, a central Data Quality Team conducts additional cross-country consistency checks and thematic reviews. The team identifies inconsistencies across indicators and countries, and may request additional review where necessary.

Methodological questions arising during data collection and review are managed through a centralised Knowledge Forum. This mechanism provides researchers and reviewers with authoritative guidance on coding decisions and promotes consistency across countries. All survey activity is recorded through an audit trail that documents comments, revisions, approvals, and requests for correction.

#### 4.5 Secondary Data Sources

The Enabling Conditions pillar is constructed using secondary indicators drawn from established international datasets and research initiatives. The selection process was guided by a combination of conceptual and methodological criteria, including: relevance to the topic; distinctiveness and non-redundancy; global coverage, availability, and comparability; credibility of sources; and availability of data.

Preference was given to indicators produced by recognised international organisations and research institutions that employ transparent methodologies and provide broad country coverage.

Secondary indicators achieve an average country coverage of over 96 per cent. Where data are unavailable for specific countries, imputation procedures are applied according to predefined rules designed to maximise coverage while maintaining cross-country comparability.

Detailed information on indicator sources, coverage, normalisation procedures, and missing-data imputation is provided in *Annexure A*.

## 5. Global Survey Methodology

### 5.1 Overview

The Global Survey is the primary data collection instrument used to construct twenty-three indicators across the AI Policy and CSO Engagement pillars, as well as the assessment of government deployment of Unacceptable Risk AI Systems (URAI). It is an evidence-based instrument: researchers are required to identify and submit publicly available, verifiable sources supporting all coding decisions, including legislation, regulations, policies, strategies, institutional documentation, government reports, consultation records, and other publicly available sources *(Table 4)*.

The Global Survey is organised around different questionnaires covering:

- AI policy frameworks;
- Government-led initiatives;
- Civil Society Engagement (CSE) initiatives;
- Government Mechanisms for CSO Inclusion in AI Policy and Governance (GMC); and
- Government deployment of AI Systems with Unacceptable Risk (URAI).

*Table 4. Scale of Data Collection*

| | |
|---|---|
| Countries and jurisdictions | 135 |
| Frameworks assessed | 376 |
| Government-led initiatives assessed | 828 |
| CSO initiatives assessed | 615 |
| Data points assessed | 68,138 |

**5.2 Evidence Types**

**5.2.1 AI Policy Frameworks**

AI policy frameworks are laws, regulations, policies, strategies, and guidelines issued by the national or federal government that address AI's implications within a specific indicator. In some cases, internationally enforceable frameworks -such as the European Union AI Act- are accepted as equivalent to national frameworks. Up to two frameworks may be assessed per indicator.

**5.2.2 Government-led Initiatives**

Government-led initiatives are programmes, projects, institutions, or activities designed, funded, or mandated by national or federal governments to advance responsible AI governance. They provide evidence of implementation of existing frameworks or, where no framework exists, of direct government action toward the indicator. Up to three initiatives may be assessed per framework (maximum two frameworks per indicator), or up to three initiatives where no framework exists.

**5.2.3 Civil Society Engagement (CSE) initiatives**

CSO Engagement indicators assess whether civil society organisations — including associations, advocacy groups, research organisations, unions, and community-based organisations — are engaged in shaping, implementing, or monitoring AI governance. One CSO Engagement questionnaire is completed per dimension, capturing up to one initiative per initiative type (six types in total).

**5.2.4 Government Mechanisms for CSO Inclusion in AI Policy and Governance**

For the fifth dimension (AI Use in Public Service), civil society engagement includes a dedicated indicator, Government Mechanisms for CSO Inclusion in AI policy and Governance (GMC), assessing whether governments have frameworks and mechanisms (foreseen and/or ongoing) that enable consultation, participation, and oversight processes for CSOs to participate in AI policy and governance processes.

Unlike the CSO Engagement indicators included in the other dimensions of the index, which assess activities undertaken by CSOs themselves, this indicator focuses on government actions that facilitate civil society participation in AI governance.

#### 5.2.5 Unacceptable Risk AI Systems (URAI)

The Global Survey also collects evidence on government deployment of AI systems that pose unacceptable risks to human rights, democratic governance, or the rule of law. URAI systems are applications that are incompatible with the obligations governments hold toward their populations. The categorisation used by GIRAI draws on the EU AI Act's classification of unacceptable risk uses and includes: biometric surveillance, criminal justice and law enforcement misuse, social scoring systems, AI-driven cyberattacks, disinformation or influence operations, AI weapons or AI use in armed conflicts, and discriminatory public service systems. Up to one case may be identified per country for each of these seven categories.

The assessment is based on the premise that governments have a particular responsibility to ensure that the use of AI in the exercise of public authority is consistent with human rights, democratic principles, and the public interest. Unlike other indicators, where governments and other actors tend to publicise their actions, URAI evidence relies on publicly documented cases or credible reports, which may be subject to availability constraints, particularly in contexts with limited civil society engagement or restricted access to information. The indicator is nonetheless essential to measuring responsible AI governance: governments' responsibilities extend beyond legislation and implementation to include refraining from using AI in ways that inflict harm on the people they serve.

Further details on the Global Survey methodology, including questionnaire instruments, indicator definitions, researcher guidance, and data collection procedures, are available at the GIRAI Public Repository.

## 6. Computing the GIRAI Score

GIRAI scores are derived from 38 indicators. 37 are organised across five equally weighted dimensions, each combining three pillars: AI Policy, CSO Engagement, and Enabling Conditions. The 38th -the Unacceptable Risk AI Systems (URAI) indicator- is applied separately as a penalty to the final country score.

Scores are computed in several steps:

Step 1. Qualitative evidence from the Global Survey is coded into numerical values, with higher values reflecting stronger responsible AI governance outcomes. Binding frameworks receive higher scores than non-binding instruments, and broader thematic coverage scores higher than partial coverage. Enabling Conditions indicators, drawn from secondary sources, are reviewed for missing values, outliers, and directionality so that higher values consistently indicate better performance.

Step 2. All indicators are then normalised to a common 0–100 scale.

Step 3. Within each dimension, normalised indicator scores are averaged into pillar scores, which are then combined as a weighted average: 60% for AI Policy, 10% for CSO Engagement, and 30% for Enabling Conditions.

Step 4. The five dimension scores are averaged with equal weight to produce the raw GIRAI score.

Step 5. Countries with documented cases of government deployment of AI systems posing unacceptable risks to human rights or democratic governance receive a proportional reduction  from 4% for one documented case to a maximum of 10% for four or more, ensuring that governance failures are not offset by stronger performance elsewhere.

$$GIRAI = GIRAI_raw \times URAI\ penalty$$

Full details of the computation — variable definitions, scoring tables, normalisation procedures, and imputation rules — are provided in the *Technical Annexures*.

**7. Scope and Limitations**

The GIRAI measures the extent to which countries are establishing governance frameworks, implementation measures, civil society engagement activities and mechanisms, and enabling conditions associated with responsible AI, providing a structured and globally comparable assessment.

As with any composite indicator, GIRAI simplifies complex phenomena in order to enable comparison across countries. While the index adopts a multidimensional approach, scores and rankings necessarily reduce complex institutional, legal, and social realities into summary measures. Results should therefore be interpreted alongside the underlying evidence and qualitative findings.

GIRAI measures the existence and characteristics of governance measures related to responsible AI, including frameworks, government-led initiatives, civil society engagement, and enabling conditions. It does not directly assess the effectiveness of these measures in achieving their intended objectives, nor does it measure the real-world impacts of AI systems on human rights, democratic governance, or social outcomes.

The index relies on publicly available and verifiable evidence collected through a structured research process. Differences across countries in transparency practices, access to information, publication requirements, and the availability of documentation may affect the amount of evidence that can be identified and validated. Although extensive quality assurance procedures were implemented, variations in evidence availability may influence results.

Finally, as with all composite indicators, GIRAI requires methodological choices regarding indicator selection, scoring, weighting, normalisation, and aggregation. To assess the implications of these choices, the Index underwent an independent statistical pre-audit and robustness assessment. While the results indicate that the overall framework is statistically coherent and robust, alternative methodological assumptions may produce some variation in country scores and rankings.

## Technical Annexures

### A. Computing the GIRAI Score

#### A.1 Overview

This chapter describes how evidence collected through the Global Survey and secondary data sources is transformed into indicator scores, pillar scores, dimension scores, and the final GIRAI score.

The GIRAI combines two types of evidence. Primary data collected through the Global Survey is used to assess the AI Policy and CSO Engagement pillars, as well as government deployment of Unacceptable Risk AI Systems (URAI). Secondary data is used to construct the Enabling Conditions pillar.

The scoring follows five main steps. First, indicator scores are computed from evidence collected through the Global Survey and from secondary datasets. Second, indicators are normalised to a common 0–100 scale. Third, indicator scores are aggregated into pillar and dimension scores. Fourth, the five dimension scores are aggregated to produce the raw GIRAI score. Finally, the URAI penalty is applied to obtain the final GIRAI score.

The following sections describe the computation of each component of the GIRAI score. For reproducibility purposes, the variable names used in the database are indicated in parentheses.

#### A.2 Computation of Indicators from Primary Data

Of the 38 indicators included in the GIRAI framework, 23 are assessed through the Global Survey. These comprise the 17 AI Policy indicators, the 5 CSO Engagement indicators, and the Unacceptable Risk AI Systems (URAI) indicator.

#### A.2.1 AI Policy Pillar

For each of the 17 AI Policy indicators listed in *Table 1*, the AI Policy score is computed as the sum of two components: Frameworks and Initiatives.

(1) Frameworks (*frameworks*, with maximum score of 16)

(2) Initiatives (*initiatives*, with maximum score of 6)

*Eq A.1* $ai_policy_raw = frameworks + initiatives$ *(for each of the 17 indicators)*

Unless otherwise specified, the equations in Sections A.2.1.1 and A.2.1.2 refer to a given country and indicator.

**A.2.1.1 Frameworks score**

For each country/indicator pair, the Frameworks score is computed from the existence of one or two policy frameworks (*fr1* and *fr2*).

If two frameworks are recorded, the score equals the mean of the framework scores. This assigns an equal weight to each framework, because they are assumed to be complementary. If only one framework is recorded, the policy framework score equals the score of the single framework (which is always *fr1*), unless its reach is partial, in which case the single framework is assigned half points. If no framework is recorded, the score is 0.

*Eq A.2* $\quad frameworks = AVERAGE(fr1, fr2) \qquad$ *if fr_counts = 2 (two frameworks)*

$\quad frameworks = fr_reach * fr1 \qquad$ *if fr_counts = 1 (one framework)*

First, for each framework applicable to each country/indicator pair, the respective score (*fr1* or *fr2*, with a maximum score of 16), is the sum of five variables:

(1) Type of document (*fr1_type, fr2_type*) follows a scoring table linked to the enforceability of the framework (*fr1_enforceability, fr2_enforceability*, Binding / Non-Binding):

| | |
|---|---|
| International Law and regulation | 12 |
| Law and regulation | 12 |
| Policy | 12 if binding, 11 otherwise |
| Strategy | 12 if binding, 11 otherwise |
| White Paper | 11 if binding, 10 otherwise |
| Guideline | 7 |
| Draft framework | 1 |
| No framework | 0 |

(2) Regulatory scope (*fr1_scope, fr2_scope*):

| | |
|---|---|
| Horizontal | 1 |
| Vertical | 0 |

(3) Stakeholder consultation (*fr1_consultation, fr2_consultation*, not collected for Guideline or Draft Framework):

| | |
|---|---|
| Yes | 1 |
| No | 0 |

(4) Operationalisation (*fr1_operation, fr2_operation*), with a maximum score of 1, is computed from four binary variables (4.1 to 4.4 are not collected for Guideline or Draft framework, and 4.2 and 4.3 are not collected for International law and regulation or Law and regulation):

(4.1) Implementation body (*fr1_body, fr2_body)*

(4.2) Implementation plan (*fr1_plan, fr2_plan*)

(4.3) Budget provisions (*fr1_budget, fr2_budget*)

(4.4) Monitoring and evaluation (*fr1_monitoring, fr2_monitoring*)

| | |
|---|---|
| Yes | 1 |
| No | 0 |

(5) Thematic coverage (*fr1_coverage, fr2_coverage*) is the average of the applicable thematic elements, which can be up to three or up to four, depending on the indicator (*elements_max*):

| | |
|---|---|
| Yes | 1 |
| Partially | 0.5 |
| No | 0 |

Second, the scores of frameworks that include exemptions for defence and security are penalised by applying a 0.9 multiplier (*fr1_defence_and_security* and *fr2_defence_and_security*).

Third, the variables *fr1_reach* and *fr2_reach* have special fields for the indicators AI Literacy and Public Sector Skills Development, and are not collected for the indicators Public Disclosure of Government Algorithmic Systems and Public Procurement, because in these two last cases applicability is limited, by definition, to state actors.

For any country/indicator pair, if there is a single framework, and it has partial reach, in computing the Frameworks component score, the score of *fr1* is multiplied by half, but there is no penalty if there are two frameworks. This implies the creation of a single indicator (*fr_reach*) which is computed only in the case of a single framework, as follows:

| Indicator | Multiplier *fr_reach* |
|---|---|
| AI Literacy | |
| All levels | 1 |
| Primary OR Secondary OR tertiary | 0.5 |
| Combination but not all | 0.5 |
| Public Sector Skills Development | |
| Both public sectors | 1 |
| Civil service only | 0.5 |
| Judiciary only | 0.5 |

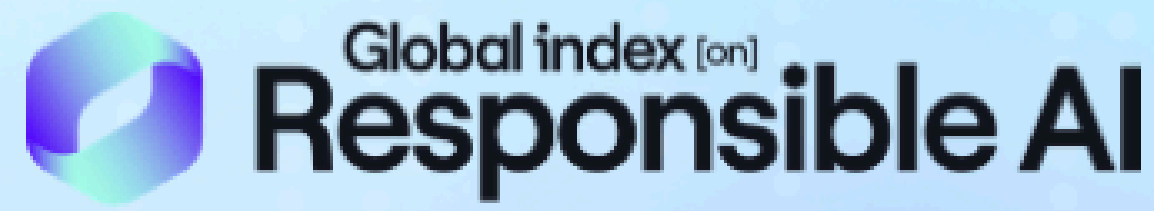
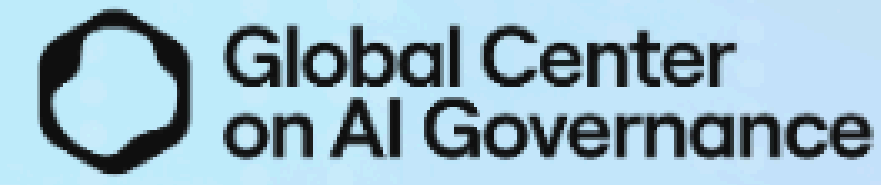

| | |
|---|---|
| Public Disclosure of Government Algorithmic Systems | 1 |
| Public Procurement | 1 |
| All other indicators | |
| Both actors | 1 |
| State only | 0.5 |
| Non-State only | 0.5 |

*Equations A.3 to A.6, specified for fr1, apply to fr2 as well:*

*Eq A.3* $fr1 = (fr1_type + fr1_scope + fr1_consultation + fr1_operation + fr1_coverage)$

$\times 0.9$ *if fr1_defence_and_security = Yes (exemption)*

*Eq A.4* $fr1_operation = n/a$ *if fr1_type = Guideline, Draft framework or No framework*

$fr1_operation = AVERAGE$

$(fr1_body, fr1_monitoring)$ *if fr1_type = International law and regulation or Law and regulation*

$fr1_operation = AVERAGE$

$(fr1_body, fr1_plan, fr1_budget, fr1_monitoring)$ *otherwise*

*Eq A.5* $fr1_coverage = SUM\ (fr1_element1, fr1_element2, fr1_element3,$

$fr1_element4$ *where applicable*$) / elements_max$

*Eq A.6* $fr_reach = 1$ *if:*

$fr_counts = 1$ *(there is one framework fr1) AND*

*Indicator = Public Disclosure of Government Algorithmic Systems, OR*

*Indicator = Public Procurement, OR*

*fr1_reach = "Both actors", "Both public sectors" or "All levels" (full reach)*

*fr_reach = 0.5 otherwise (i.e. when fr1_reach = partial reach)*

**A.2.1.2 Initiatives score**

For each country/indicator pair, the Initiatives score is computed from the existence of two sets (*init1* and *init2*) of up to three policy initiatives associated with the implementation of each of the two policy frameworks, for a total of up to six initiatives per indicator.

Policy initiatives may also exist in absence of a framework, in which case they are interpreted as government actions towards the indicator and listed under *init1*. In those cases, only one set of initiatives is recorded.

Note that for a given country/indicator pair, the second framework *fr2* might have implementation initiatives (*init2_counts > 0*), even as the first, *fr1,* might not (*init1_counts* = 0).

The Initiatives score, with a maximum of 6, is obtained as twice the maximum count of initiatives per framework *(init1_counts* and *init2_counts*, from 0 to 3).

*Eq A.7* *initiatives = MAX (init1_counts, init2_counts) * 2*

**A.2.2 CSO Engagement pillar**

In GIRAI's conceptual framework, the CSO Engagement pillar is concerned with five indicators, one for each dimension *(Table 1)*, with scores ranging from 0 to 1.

1. Civil Society Engagement in Inclusion and Diversity
2. Civil Society Engagement in Ethics and Sustainability
3. Civil Society Engagement in Labour and Skills
4. Civil Society Engagement in Trust and Safety
5. Government Mechanisms for CSO Inclusion in AI Policy and Governance

*Eq A.8* *cso_engagement_raw = {CSE scores for dimensions 1 to 4, GMC score for dimension 5}*

**A.2.2.1 CSE scores**

Initiatives led by Civil Society Organizations (CSO-led initiatives) apply to the first four dimensions: Inclusion and Diversity; Ethics and Sustainability; Labour and Skills; and Trust and Safety, with one indicator per dimension.

For each country/dimension pair, researchers may record up to six CSO-led initiatives. Each Civil Society Engagement (CSE) score captures both the existence of initiatives and their respective contributions to AI Policy indicators. The score is expressed on a (0–1) scale and is computed in two steps:

First, for each of the up to six recorded CSO-led initiatives, the contribution score (*cse1 to cse6*) is calculated as a proportion, the number of indicators to which the initiative contributes, over the total number of indicators within the respective dimension (*contributions_max*), therefore ranging from 0 to 1.

These initiatives can contribute to the 14 AI Policy indicators included in the first four dimensions and measured in the Global Survey, plus four additional contribution areas used only for CSE-denominator purposes: Rights of Persons with Disabilities (in D1), and Data Protection and Privacy, Data Sharing and Access, and Consumer Protection (D4). Rights of Persons with Disabilities is not an Enabling Conditions indicator, but is included in the CSE contribution denominator for Inclusion and Diversity.

Therefore, the total number of indicators considered in each dimension is:

- Inclusion and Diversity: four indicators
- Ethics and Sustainability: four indicators
- Labour and Skills: three indicators
- Trust and Safety: seven indicators.

For example, in the second dimension, an initiative contributing to three indicators, where the total number of indicators considered is four, receives a score of 3/4 = 0.75.

Second, for each dimension, the CSE score (*cse*) is calculated as the average positive contributions across recorded CSO-led initiatives, with an adjustment (*cse_adjustment*) that accounts for the number of initiatives recorded. This adjustment applies to countries with three or less initiatives:

| Number of recorded initiatives | CSE adjustment |
|---|---|
| 1 | 0.25 |
| 2 | 0.50 |
| 3 | 0.75 |
| 4 to 6 | 1.00 (no adjustment) |

*Eq A.9.1* *cse1 to cse6 = contributions / total number of indicators in the respective dimension*

*Eq A.9.2* *cse = AVERAGE (cse1 to cse6 > 0) × cse_adjustment*

**A.2.2.2 GMC score**

In the fifth dimension, AI Use in Public Service, CSO engagement is captured through a single indicator that does not follow the methodology described above: Government Mechanisms for CSO Inclusion in AI Policy and Governance.

This indicator examines whether governments have established mechanisms through which CSOs participate in AI governance across three complementary aspects: consultation in policy drafting (*gmc_consultation*); provisions for ongoing participation or oversight, including their thematic coverage (*gmc_policy_provisions*); and existing and operational mechanisms currently functioning in practice (*gmc_mechanisms*). When the national AI policy or equivalent framework is a draft, the policy provisions component is multiplied by 0.5 (*gmc_draft*) to reflect that draft instruments are not yet fully adopted.

For each country, the GMC score is expressed on a (0–1) scale and is computed as a weighted sum of these three aspects:

*Eq A.10* *gmc = 0.3 × gmc_consultation + 0.3 × gmc_policy_provisions + 0.4 × gmc_mechanisms*

*where*

*gmc_consultation = 1 if gmc_consult_counts > 0, 0 otherwise*

*gmc_policy_provisions = AVERAGE (gmc_provisions, gmc_coverage) * gmc_draft*

*gmc_provisions = 1 if gmc_particip_provis_counts > 0, 0 otherwise*

*gmc_coverage = AVERAGE (gmc_element1, gmc_element2, gmc_element3)*

*gmc_draft = 0.5 if gmc_type = Draft framework, 1 otherwise*

*gmc_mechanisms = 1 if gmc_particip_mech_counts > 0, 0 otherwise*

The thematic coverage (*gmc_coverage*) is the average of three elements (*gmc_element1 to gmc_element3*), covering Diversity and inclusion, Accountability, and Availability of resources, respectively:

| | |
|---|---|
| Yes | 1 |
| Partially | 0.5 |
| No | 0 |

**A.2.3 Unacceptable Risk AI score (URAI penalty)**

The 18th indicator, Unacceptable Risks Artificial Intelligence Systems (URAI), is assessed through a distinct questionnaire and computed separately as a penalty applied to the final country GIRAI score.

This measure concerns 35 countries (out of 135) and captures documented cases of government deployment of AI systems posing unacceptable risks to rights or democratic processes, i.e. of misuse of AI in public service. It therefore reflects failures in AI governance and implementation.

The penalty follows a threshold-based approach. The calculated variable *urai_counts* records the number of documented deployments, which can be up to seven. In practice, no country in the current dataset has more than four recorded deployments; therefore, the database displays fields only up to *urai4*. The count is then recoded into *urai_penalty*, expressed as a proportional reduction applied to the final country score:

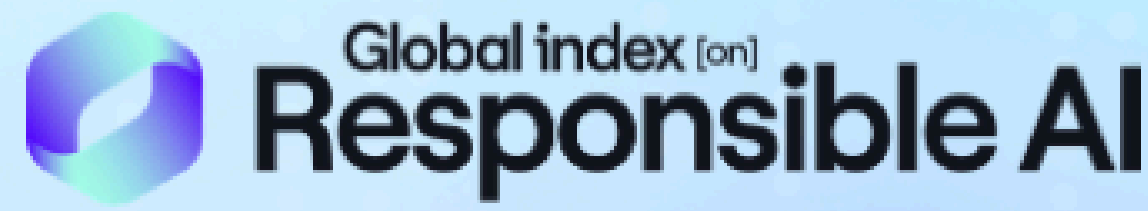

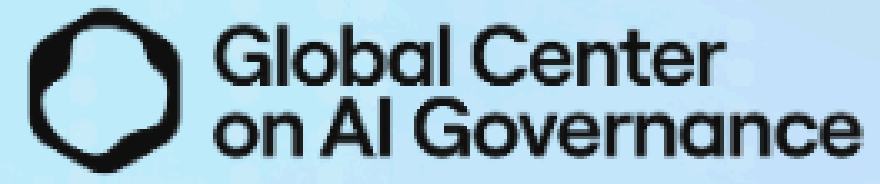

| *urai_counts* | *urai_penalty* |
|---|---|
| 0 | 1.00 |
| 1 | 0.96 |
| 2 | 0.93 |
| 3 | 0.91 |
| 4 or more | 0.90 |

### A.3 Computation of indicators from secondary data

A total of 15 indicators derived from secondary sources are used to construct the Enabling Conditions pillar *(Table 5)*. These indicators are drawn from internationally comparable datasets and incorporated into the GIRAI framework.

*Table 5: Secondary-source indicators included in the Enabling Conditions pillar*

| GIRAI Dimension | GIRAI Indicator | Source Indicator | Source | Country coverage |
|---|---|---|---|---|
| Inclusion and Diversity | Egalitarian Democracy | Egalitarian Democracy Index | Varieties of Democracy (V-Dem) | 132/135 |
| | Device affordability | Device affordability for lowest 40% | Mobile Connectivity Index | 131/135 |
| | Gender gap in mobile internet | Gender gap in mobile internet | Mobile Connectivity Index | 131/135 |
| Ethics and Sustainability | Rule of Law | Rule of Law | Worldwide Governance Indicators (WBGI) | 135/135 |
| | Low-Carbon Energy Share | Share of electricity generated by low-carbon sources | Our World in Data | 131/135 |
| Labour and Skills | Skills and Literacy | Skills and Literacy | Digital Development Compass | 132/135 |
| | Labour Rights Compliance | Fundamental labour rights are effectively guaranteed | World Justice Project (WJP) | 116/135 |

| GIRAI Dimension | GIRAI Indicator | Source Indicator | Source | Country coverage |
|---|---|---|---|---|
| Trust and Safety | Cybersecurity | Global Cybersecurity Index | International Telecommunication Union (ITU) | 133/135 |
| | Data Protection and Privacy | Data and Privacy | Digital Development Compass | 133/135 |
| | Data Sharing and Access | Combination of Status of Implementation, Sectoral scope, Permitted Participants in Data Exchange Systems | DPI Map - Data Exchange | 135/135 |
| | Consumer Protection | Consumer Protection | Digital Development Compass | 132/135 |
| | Global Peace | Global Peace Index | Vision of Humanity / Institute for Economics & Peace | 130/135 |
| AI Use in Public Service | Civil Society Accountability | Diagonal Accountability Index | Varieties of Democracy (V-Dem) | 132/135 |
| | Public Service Delivery | Public Service Delivery Index | World Bank GovTech Dataset | 133/135 |
| | Access to Public Information | Combination of 3.2 Right to Information and 3.1 Publicized laws and government data | World Justice Project (WJP) | 116/135 |

The rationale details for each secondary data source including the rationale for their inclusion, the source definition and notebooks with the data processing are available at GIRAI Public Repository.

### A.3.1 Composite indicators included in the Enabling Conditions pillar

Two Enabling Conditions indicators were constructed by combining multiple source variables into composite measures. This was done where no single source variable fully captured the concept measured by the GIRAI indicator, or where combining related variables provided a more complete measure.

Data Sharing and Access indicator: a composite score was constructed by combining three variables from the DPI Map Data Exchange Systems (DES) dataset: Status of implementation, Sector-specific / Cross-sectoral scope, and Permitted participants. These three variables were selected based on their conceptual relevance and their relatively higher coverage compared to other DES variables. Scores were assigned through a lookup table covering all possible combinations, running from 0 (no system reported) to 16 (system fully implemented, cross-sectoral, and open to public, civil society, and private sector participants). Status of implementation carries the greatest weight, followed by scope, and then permitted participants.

Within each combination, missing values in Permitted participants are treated conservatively: they receive the same score as "Public only", on the grounds that absence of documentation should not be penalized beyond what is already captured by the other two variables. The resulting scale has 13 realized values across the sample, with gaps arising from the decision to treat Public and NA as equivalent in the participants' dimension, as shown in *Table6*.

*Table 6. Scoring Framework for the Data Sharing and Access Composite Indicator*

| | | Permitted participants | | |
|---|---|---|---|---|
| **Status** | **Scope** | **Public / NA** | **Public + one additional stakeholder group** | **Public, Civil, Private** |
| Implemented | Cross-sectoral | 14 | 15 | 16 |
| Implemented | Sector-specific / NA | 10 | 11 | 12 |
| Planned/Piloted | Cross-sectoral | 6 | 7 | 8 |
| Planned/Piloted | Sector-specific / NA | 2 | 3 | 4 |
| NA | Any | 0 | 0 | 0 |

For the indicator Access to Public Information indicator:, a composite score was constructed by combining two World Justice Project variables: 3.1 Publicized laws and government data and 3.2 Right to information. These variables are conceptually closely related and show a high positive correlation (r = +0.813), supporting the use of a simple arithmetic mean. The indicator measures the extent to which the legal and institutional framework supports access to public information, capturing both the public availability of laws and government data and the effective implementation of right-to-information mechanisms. The resulting score is a continuous index ranging from 0 to 1, where higher values indicate stronger access to public information.

**A.4 Imputation of missing data**

Average country coverage across these 15 indicators is 96.4%, with a median of 97.8%; two indicators have full coverage, and five cover at least 133 of the 135 target countries.

Missing values are concentrated in a relatively small number of country and territory cases: Palestine (8 indicators), Kosovo (7), Antigua and Barbuda (5), Hong Kong SAR (5), Turkmenistan (4), and Belize, Saint Lucia, the Central African Republic and Lesotho (3 each), suggesting that missingness is driven primarily by country coverage limitations.

Before imputation, the original source of each affected indicator was checked for earlier reference years that could support same-source imputation; where available, such historical data were used in preference to other methods.

As a general rule, missing values are imputed using peer-group means based on the UNSD[16] intermediate region and World Bank income group as the default method. Where no group of countries could be identified, the hierarchy falls back successively to the UNSD subregion and, if necessary, to the broader UNSD region, combining them in each case with the same income group. Where more than one plausible comparator group was available, sense checks were conducted to verify that the imputed values remained within a plausible range and did not materially distort country scores. The selected donor groups may therefore vary by indicator, depending on the most appropriate comparator set for each variable.

**A.5 Treatment of outliers**

Outlier treatment is triggered when an indicator exhibits a statistically problematic distribution, defined by an absolute skewness greater than 2 and a Pearson kurtosis greater than 3.5. None of the 15 indicators exceeded these thresholds; therefore, no outlier treatment was applied.

**A.6 Computation of the GIRAI Index**

GIRAI indicators are derived from two sources: primary data collected through the Global Survey and secondary data from internationally comparable datasets. This section outlines the computation steps applied to both types of indicators.

**A.6.1 Normalization of indicators**

Indicator values are normalized to a [0, 100] scale prior to aggregation. AI Policy and GMC indicators derived from the Global Survey use a special min-max normalization based on theoretical scoring bounds: [0, 22] for AI Policy and [0, 1] for GMC. CSE indicators are normalized by the joint sample maximum [0, 0.8125] across the four CSE indicators, reflecting their common computation logic and preserving comparability across CSE indicators.

[16] Kosovo is not UN Member States; it is classified here for statistical purposes only.

*Eq A.11* *ai_policy = ai_policy_raw / 22 * 100* *(for each of the 17 indicators)*

*Eq A.12* *cso_engagement = {CSE scores * 81 for dimensions 1 to 4, GMC score *100 for dimension 5}*

Indicators derived from secondary sources follow sample min-max normalization, in line with standard composite indicator practice.

*Eq A.13* *score = (value − min) / (max − min) × 100 for a good*

*score = 100 - (value − min) / (max − min) × 100 for a bad*

Secondary-source indicator Global Peace is a "bad", meaning that higher raw values indicate lower performance.

**A.6.2 Aggregation of indicators at the country level**

The GIRAI being a matrix, scores can be computed in at least two ways. From a policy perspective, pillar scores are interesting, as they relate to different stakeholders, such as AI policy-makers for AI Policy (the first pillar); the ecosystem of civil society and CSOs engaged in advocacy on IA issues for CSO engagement (the second pillar); all evolving, together with the general public, in a context determined by relevant enabling conditions (the third pillar). From the perspective of particular areas of interest, however, scores at the dimension level might be the relevant indicator.

Dimension scores are computed in two steps. First, within each pillar and each dimension (cells in *Table 1*), dimension/pillar scores are obtained as the simple arithmetic mean of the normalized indicator scores. All indicators carry equal weight, even though the number of indicators varies across dimensions.

*Eq A.14* *dimension/pillar score = AVERAGE (indicator scores within the pillar and dimension)*

*(computation of ai_policy_d, cso_engagement_d and enabling_conditions_d, d from 1 to 5)*

The weighting scheme builds on both internal methodological discussions and continuity with the previous edition of the GIRAI. In the first edition, pillar weights were derived through a structured "budget allocation exercise" conducted with the GIRAI Core Team, where participants distributed 100 points across pillars based on criteria such as relevance for responsible AI outcomes, policy relevance, and data reliability and quality. This process resulted in a weighting structure where frameworks and government actions together accounted for 80% of the total score, reflecting their central role in shaping and implementing AI governance, while non-state actors accounted for 20%.

The current weighting maintains this underlying logic while adapting it to the revised structure of the index. The AI Policy pillar (60%) continues to capture the core regulatory and policy initiatives of governments, consolidating what was previously split between frameworks and government actions, and therefore retains the largest share of the weight. The CSO Engagement pillar (10%) captures the role of civil society in shaping, monitoring, and complementing governance efforts; while essential, it is assigned a lower weight to reflect both its indirect influence on formal policy outcomes and greater variability in data availability and comparability across countries. Finally, the Enabling Conditions pillar (30%) reflects the importance of structural factors that shape the feasibility and effectiveness of responsible AI governance.

Overall, the weighting preserves the core assumption from the first edition: that government-led policy and regulatory action is the primary driver of responsible AI outcomes, while civil society engagement and contextual conditions play supporting roles.

Therefore, for each of the five dimensions (rows in *Table1*), country dimension scores are computed as the weighted arithmetic mean of the three respective dimension/pillar scores (0.6 for AI Policy, 0.1 for CSO Engagement, and 0.3 for Enabling Conditions).

*Eq A.15* *dimension score = 0.6 * ai_policy_d + 0.1 * cso_engagement_d + 0.3 * enabling_conditions_d*

For descriptive and policy analysis purposes, country-level pillar scores (*ai_policy, cso_engagement and enabling conditions,* columns in *Table 1*) may also be calculated as the simple arithmetic mean of the corresponding dimension/pillar scores across the five dimensions.

*Eq A.16* *pillar scores = AVERAGE (respective dimension/pillar scores across the five dimensions)*

*(computation of ai_policy, cso_engagement and enabling_conditions)*

The raw GIRAI index score can then be computed in two ways, as the simple arithmetic mean of the five dimension scores, or as the weighted arithmetic average of the pillar scores.

*Eq A.17* *GIRAI_raw = AVERAGE (five dimension scores)*
*GIRAI_raw = 0.6 × ai_policy + 0.1 × cso_engagement + 0.3 × enabling_conditions*

The final GIRAI score is then obtained by applying the URAI penalty to the raw GIRAI score.

*Eq A.18* *GIRAI = GIRAI_raw × urai_penalty*

The penalty is applied after the raw GIRAI score has been computed from the five dimension scores, as a separate final adjustment. This makes the penalty transparent, avoids altering the interpretation of indicator, pillar, and dimension scores, and ensures that documented misuse by governments is not fully offset by stronger performance elsewhere in the index.

The effect of the penalty scales proportionally with the country's raw GIRAI score, at a decreasing rate. For example, if a country deploys one count of URAI system, the adjusted score is 0.96 the raw GIRAI score; with two counts, it is 0.93; with three counts, it is 0.91; and with four or more counts, it is 0.9.

## B. Statistical Pre-Audit and Robustness Assessment

### B.1 Purpose and scope of the Pre-Audit

To assess the robustness of GIRAI results, the Index underwent an independent statistical pre-audit conducted using an adapted version of the COIN tool, which is a statistical auditing tool developed by the Joint Research Centre (JRC) Competence Centre on Composite Indicators (COIN)[17].

The objective of the pre-audit was to evaluate the extent to which GIRAI scores and rankings remain stable under alternative and plausible methodological assumptions. The pre-audit examined the conceptual and statistical coherence of the framework, the sensitivity of country scores and rankings to alternative modelling assumptions, and the influence of individual indicators on overall results.

### B.2 Missing data and imputation

The pre-audit compared the index using imputed data and zero-imputation as a robustness check. For most countries, the impact of imputation on rankings was limited. However, a small number of countries showed larger rank shifts, particularly Antigua and Barbuda and the State of Palestine, which shifted by 10 or more positions, and Hong Kong SAR; Oman; and Ukraine, which shifted by four or more positions. For Antigua and Barbuda and Palestine, the pre-audit notes that it cannot be ruled out entirely that the imputation method may influence results rather than the countries' actual policy conditions.

No imputation, however, is probably less optimal. In effect, in any index, non-imputation implies that missing data is de facto imputed as the weighted mean of indicators of the same country within the same dimension/pillar, which arguably is less optimal than imputation by peer-group for the same indicator, in attention to topic, scale, and range.

### B.3 Conceptual and Statistical Coherence of the Framework

It is expected and desirable for the overall robustness of the index that indicators and dimensions/pillars be mostly positively – but not strongly – correlated and statistically balanced.

The analysis performed on the GIRAI components shows that:

- there is no case of strong collinearity (i.e. Pearson correlation coefficients greater than 0.92);

---

[17] Becker, W., Benavente, D., Dominguez Torreiro, M., Moura, C., Neves, A., Saisana, M. and Vertesy, D., COIN Tool User Guide (Publications Office of the European Union, 2019) ISBN 978-92-76-12385-9, doi:10.2760/523877.

- at the indicator level, only two indicators present correlations below the 0.30 suggested threshold: Low-Carbon Energy Share and Rule of Law (0.14); Civil Society Accountability and Public Service Delivery (0.23);
- at the dimension level, there is a strong correlation of the indicator Fairness and Non-Discrimination (0.93) with the upper dimension. The framework could have been rebalanced with a lower weight for this indicator, as it is somewhat redundant, against a higher weight for Environmental impact (correlation of 0.70 with the upper dimension), but the team decided to keep the equal weights for parsimony and communications purposes.

### B.4 Impact of Modelling Assumptions and Robustness of Results

Scores and rankings depend on modelling choices: the pillar and dimension structure, the treatment of indicators, imputation of missing data, normalisation and bounds used, weights, and aggregation methods. These choices are based on expert opinion, common practice, statistical analysis, or simplicity.

The robustness analysis assesses to what extent these choices might affect rankings. This analysis is based on a multi-modelling approach to calculate scores and rankings under conditions of uncertainty[18].

#### B.4.1 Modelling choices

The modelling variations in the JRC auditing tool result in 48 combinations ((5+2)×3×2+2+2+1+1=48). Of these, 47 are used to define confidence intervals for rankings; the 48th — rankings computed on raw (non-imputed) data — is included in the tool for reference and comparison purposes only. The 47 sets comprise:

- Five (5) linear normalization methods: GIRAI; min-max normalisation; data max (i.e. min set at zero); goalposts (i.e. original indicator theoretical bounds for all indicators); and Z-scores (standardised scores, with an average of 50 and a standard deviation of 10 so that the minimum and maximum scores are within the 0-100 range).

- Two (2) non-linear normalization methods: median-min-max (transformation so that the median value of positive scores is 50 for all indicators); and data percent ranks.

- Each of these seven methods included the following:

---

[18] Saisana, M., Saltelli, A. and Tarantola, S., 'Uncertainty and Sensitivity Analysis Techniques as Tools for the Analysis and Validation of Composite Indicators' (2005) 168(2) Journal of the Royal Statistical Society A 307; Saisana, M., D'Hombres, B. and Saltelli, A., 'Rickety Numbers: Volatility of University Rankings and Policy Implications' (2011) 40 Research Policy 165.

o Three sets of weights for each method (7x3): GIRAI weights, equal weights, and random weights (weights defined randomly within a pre-specified range).

o Two types of aggregation at the pillar level (7x3x2) : arithmetic and geometric averages (maintaining the arithmetic average at lower levels). For geometric averages, to handle the presence of zeroes and moderate the non-compensatory penalty on highly unbalanced profiles, pillar scores are shifted upward by adding 50, and 50 is subsequently subtracted from the geometric GIRAI raw score to restore the 0-100 scale.

• Four (4) sets of rankings based on ranks (the conceptual framework is not considered; indicators are directly aggregated into a ranking or an index):

o Rankings based on median ranks (rank of the median of indicator ranks).

o Rankings based on average ranks (rank of the average of indicator ranks).

o Two sets of Borda rankings: with GIRAI weights and with equal weights. The Borda method is an alternative way of aggregating indicators, based on ranks. For N countries (in this case N=135), for each indicator the top-ranked country gets N-1 points, the second-ranked country gets N-2 points and so on, and the last-ranked country gets 0 points. The country's overall score is the sum (simple or weighted) of indicator points. Countries are then ranked by their overall score.

• Rankings based on the Copeland rule (1). The Copeland rule is based on the outranking matrix at the pillar level. Pillar scores are compared pairwise. Country XX is assigned a score equivalent to the sum of the weights of the pillars where it has a higher value than country YY (for example, in the first two pillars, 0.7), Country YY then gets 0.3 (they must add up to 1). Under 'absolute dominance', i.e. when a country scores 1 pairwise, methodological choices cannot affect the relative standing of XX with respect to YY in the ranking. The greater the dominance, the more robust country ranks are to methodological assumptions.

• For the sake of comparison, in case the raw data is used (instead of the imputed data), the file also includes the rankings with imputed data (1) .

Linear normalization methods (including z-scores) have the particularity that by plotting raw data against normalized data, the result is a straight line. In addition, aggregation based on these data series respects the structure and weights of the GIRAI index.

**B.4.2 Uncertainty analysis results**

Results presented here correspond to imputed data. In addition, the analysis is based on the final GIRAI rankings once the URAI penalty is applied.

The 47 sets of rankings are summarized in a series of different rankings, for which a median rank is computed, as well as an interval of ranks, for each country.

GIRAI ranks are particularly robust at the top and at the bottom, which is quite common for composite indicators, and a few countries show relatively large intervals of ranks. Considering linear models only, 32,6% of countries shift 5 positions or less, and 65,2% shift 10 positions or less.

For reporting purposes, ranking intervals are presented using a subset of 13 linear arithmetic models, with exclusion of the lowest and highest ranks per country *(Figure 2)*, as this provides a more interpretable representation of uncertainty while remaining aligned with the methodological choices adopted by GIRAI, and for the reasons explained in the sensitivity analysis *(Section B.4.3)*. The red line represents the GIRAI ranking, and, for each country, error bars represent the interval of rankings under the models considered, and the blue dots represent the median ranks.

*Figure 2: Robustness of GIRAI ranks, imputed data (13 of 15 models)*

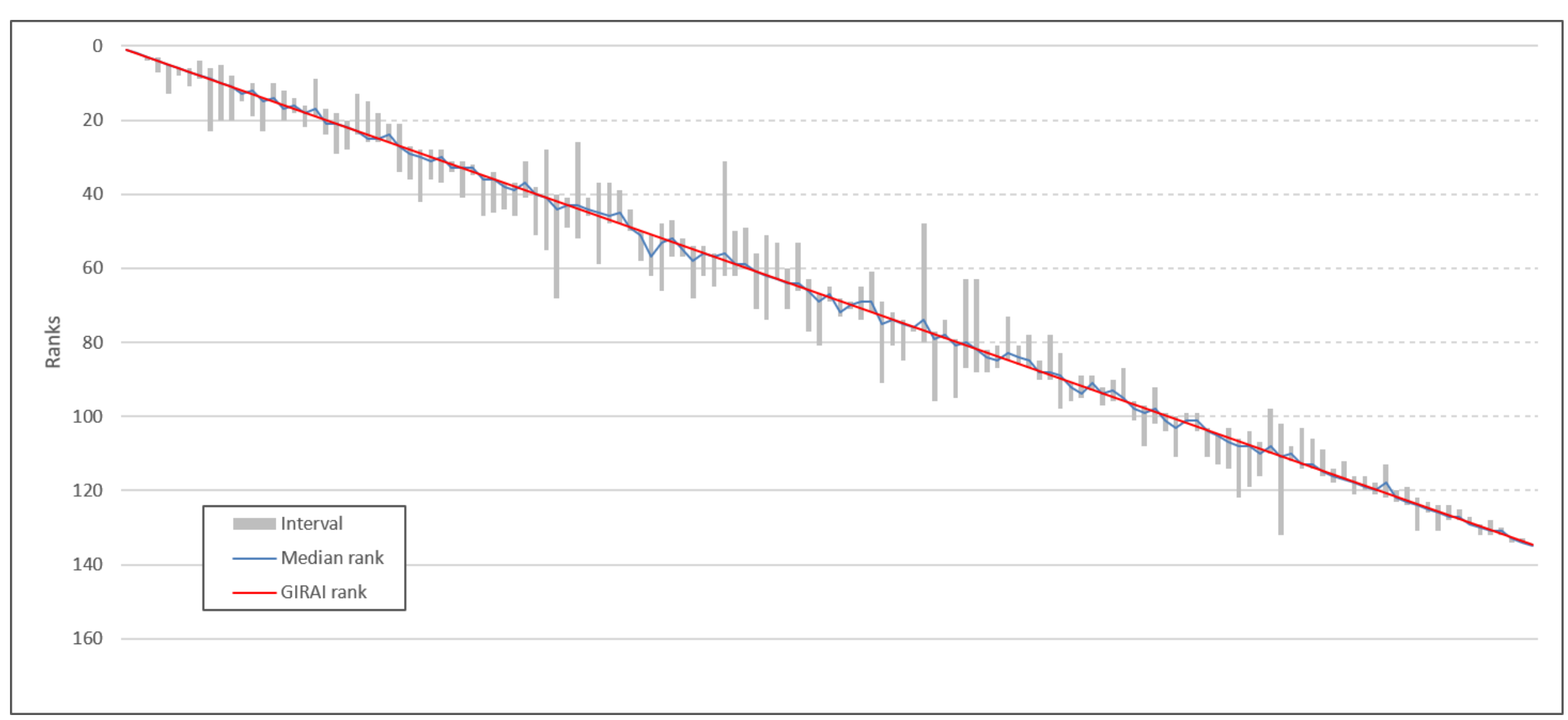

**B.4.3 Sensitivity analysis results**

Complementary to the uncertainty analysis, sensitivity analysis is used to identify which of the modelling assumptions have the highest impact on certain country ranks[19].

Results show that the sensitivity of the rankings escalates from low to high as the methodology shifts from standard linear normalization methods and arithmetic scores, to weighting schemes (GIRAI v. random weights, v. equal weights, in that order), to ordinal techniques (dataranks and Borda), and finally to non-linear techniques, with Copeland and geometric averages representing the most significant departures from the baseline (GIRAI ranks).

- The low sensitivity sets are those using arithmetic averages. This is expected for linear models and percent ranks, because GIRAI uses a slightly modified min-max normalisation method and arithmetic aggregation. But it is also the case for median minmax (non-linear) and z-scores (linear), which both center indicators (at a median of 50 and at a mean of 50, respectively), but are differently sensitive to the shape of indicators' distributions (for high-kurtosis indicators, z-scores reward low value scores, while median min-max rewards mid-values).

- Rankings based on ranks also present sensitivity. For ranks based on indicator median and average ranks, this is mostly because these models do not consider the structure and weights of the index. Borda & Copeland are sensitive because they discard the magnitude of the difference between countries and only look at the ordinal win/loss.

- High sensitivity is also associated with weighting schemes, GIRAI vs. random vs. equal (in that order, because by construction random weights are closer to GIRAI weights). The ranking is less robust to equal weights because GIRAI assigns a relatively much higher weight to AI Policy compared to CSO Engagement (0.6 against 0.1); in this case, equal weights act as a stress test with significant impact in a handful of countries, particularly in combination with geometric averages.

- Geometric averages have maximum sensitivity, and it is the modelling choice that most aggressively challenges the robustness of results, particularly in conjunction with equal weights. The main impact comes from the multiplicative nature and the handling of zeros. Arithmetic averages are fully compensatory, a low score in one indicator can be compensated by a high score elsewhere. Geometric averages, in contrast, are only partially compensatory, so that balanced profiles are rewarded with relatively higher scores. The impact is strong due to the high number of zeros in Global Survey indicators, which carry through pillars and dimensions. A secondary impact comes from the many countries with unbalanced profiles. This is mostly due to the high presence of zero or low scores, which is expected as GIRAI addresses a multidimensional issue, responsible AI, that has been under development only in very recent years and that has not been fully consolidated in terms of policy implementation and enforcement.

[19] Saltelli, A., Ratto, M., Andres, T., Campolongo, F., Cariboni, J., Gatelli, D., Saisana, M. and Tarantola, S., Global Sensitivity Analysis: The Primer (John Wiley & Sons 2008).

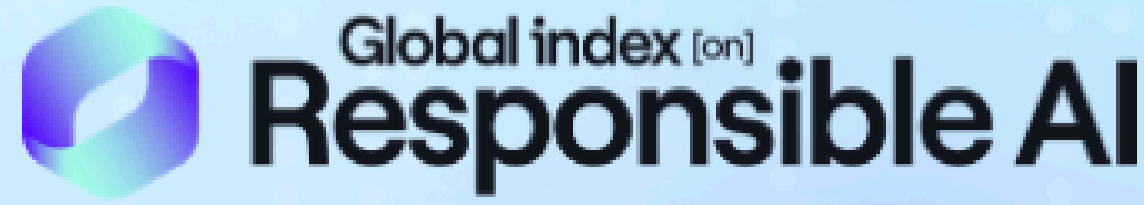

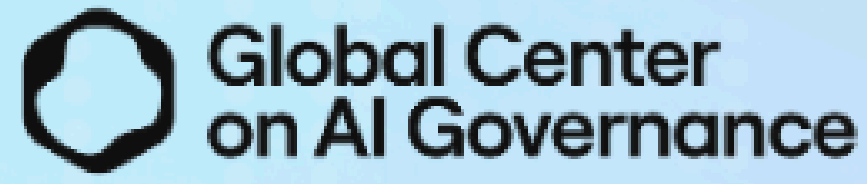

For the reasons exposed, confidence intervals are based on the linear models only (13 out of 15, after trimming extreme results). This reporting approach provides a transparent representation of ranking uncertainty while remaining consistent with the methodological assumptions adopted by GIRAI.

*Table 7: Robustness of GIRAI ranks, imputed data (13/15 linear models).*

| Country | Ranking | Interval | Country | Ranking | Interval | Country | Ranking | Interval |
|---|---|---|---|---|---|---|---|---|
| NOR | 1 | [1,1] | IND | 46 | [37,55] | BTN | 91 | [91,96] |
| ITA | 2 | [2,3] | VNM | 47 | [37,48] | BWA | 92 | [89,95] |
| IRL | 3 | [2,4] | EGY | 48 | [39,48] | TZA | 93 | [89,93] |
| FRA | 4 | [3,6] | JOR | 49 | [45,50] | LSO | 94 | [92,97] |
| NLD | 5 | [5,13] | KEN | 50 | [50,59] | ARM | 95 | [90,96] |
| DEU | 6 | [6,8] | PAK | 51 | [50,71] | ECU | 96 | [87,96] |
| GBR | 7 | [6,11] | QAT | 52 | [50,66] | MNG | 97 | [97,101] |
| SVN | 8 | [4,9] | IDN | 53 | [47,56] | PSE | 98 | [97,108] |
| LVA | 9 | [8,23] | MYS | 54 | [52,57] | BRB | 99 | [92,102] |
| EST | 10 | [5,20] | ARE | 55 | [55,69] | MWI | 100 | [99,104] |
| BRA | 11 | [8,20] | GHA | 56 | [54,62] | BGD | 101 | [99,114] |
| ESP | 12 | [12,15] | NPL | 57 | [56,66] | AGO | 102 | [99,102] |
| GRC | 13 | [10,19] | MKD | 58 | [31,62] | XKX | 103 | [99,104] |
| CHL | 14 | [14,23] | BEN | 59 | [53,62] | TJK | 104 | [103,111] |
| BGR | 15 | [10,15] | THA | 60 | [49,60] | BLZ | 105 | [105,113] |
| BEL | 16 | [12,20] | SLV | 61 | [59,71] | ATG | 106 | [103,114] |
| POL | 17 | [14,18] | OMN | 62 | [58,74] | UGA | 107 | [106,122] |
| PRT | 18 | [16,22] | MAR | 63 | [54,63] | BOL | 108 | [104,117] |
| CHE | 19 | [7,19] | CIV | 64 | [61,71] | LCA | 109 | [106,116] |
| AUT | 20 | [16,24] | ARG | 65 | [51,65] | TUN | 110 | [98,110] |
| LTU | 21 | [18,29] | RWA | 66 | [65,77] | BLR | 111 | [102,131] |
| URY | 22 | [20,28] | SAU | 67 | [67,81] | HND | 112 | [108,112] |
| AUS | 23 | [13,24] | PHL | 68 | [65,68] | ZWE | 113 | [103,114] |
| JPN | 24 | [15,26] | GEO | 69 | [67,73] | SLE | 114 | [106,114] |
| CAN | 25 | [18,26] | KAZ | 70 | [69,71] | LBN | 115 | [109,115] |
| COL | 26 | [21,26] | ALB | 71 | [63,74] | LAO | 116 | [114,118] |
| USA | 27 | [21,33] | LKA | 72 | [61,72] | TGO | 117 | [112,117] |
| CRI | 28 | [27,36] | ETH | 73 | [69,91] | GIN | 118 | [115,121] |
| HUN | 29 | [29,42] | SEN | 74 | [72,81] | CMR | 119 | [116,121] |
| ROU | 30 | [28,36] | BHR | 75 | [75,85] | MOZ | 120 | [118,121] |
| KGZ | 31 | [28,37] | AZE | 76 | [75,77] | BFA | 121 | [113,121] |
| SVK | 32 | [30,34] | ZAF | 77 | [48,79] | LBR | 122 | [120,123] |
| SRB | 33 | [31,41] | LBY | 78 | [77,96] | DZA | 123 | [119,124] |
| PER | 34 | [32,35] | ZMB | 79 | [74,79] | SOM | 124 | [122,131] |
| HRV | 35 | [35,46] | GMB | 80 | [80,95] | NER | 125 | [123,126] |
| DOM | 36 | [35,45] | MEX | 81 | [63,85] | MMR | 126 | [124,132] |
| KOR | 37 | [37,44] | MNE | 82 | [63,83] | TKM | 127 | [124,128] |
| NGA | 38 | [36,46] | UZB | 83 | [82,88] | COD | 128 | [125,128] |
| NZL | 39 | [30,41] | MUS | 84 | [81,87] | CAF | 129 | [127,129] |
| SGP | 40 | [38,51] | PRY | 85 | [73,85] | HTI | 130 | [129,132] |
| ISR | 41 | [28,52] | TTO | 86 | [81,86] | COG | 131 | [128,132] |
| CHN | 42 | [40,72] | PAN | 87 | [78,88] | BDI | 132 | [130,132] |
| HKG | 43 | [41,49] | JAM | 88 | [84,90] | TCD | 133 | [133,134] |
| UKR | 44 | [26,49] | KHM | 89 | [78,90] | AFG | 134 | [133,134] |
| MDA | 45 | [41,46] | KWT | 90 | [83,98] | SSD | 135 | [135,135] |

**B.4.4 Sensitivity to indicators**

Sensitivity analysis also provides insights into what is affecting scores the most for each country. The tool does this for the main model only, by providing the changes in ranks due to leaving out one indicator at a time. Indicators with the most impact on ranks are Low-carbon energy share (combined sum of absolute shifts in rankings of 34 positions); Cultural and Linguistic Diversity (combined shift of 30 positions); and Access to Redress and Remedy (combined shift of 24 positions). In attention to the lower weight of CSO Engagement, leaving out any of its five indicators has a lower impact, the main one being that of GMC.

### B.5 Concluding remarks

The pre-audit using the COIN tool confirms that the index is reliable and based on a statistically coherent conceptual framework. The pre-audit also recognizes the effort of developers to obtain replicable and transparent results.

With few exceptions, GIRAI is statistically balanced in its pillars, and most of the indicators provide meaningful information on the variation of scores. The sensitivity of GIRAI rankings to weighting schemes (equal weights as opposed to GIRAI pillar weights of 0.6, 0.1, and 0.3), the index structure (the GIRAI matrix in contrast with rankings based on indicators only), and the aggregation method (geometric averages penalizing unbalance and zeroes, in contrast with the compensatory nature of GIRAI's arithmetic aggregation) is important, however, impacting the robustness of results and confidence intervals for a few countries, particularly in mid-level ranks.

But this should not be perceived as a weakness of the index. On the contrary, these results underscore the importance of strengthening the data collection methodologies, particularly regarding CSO Engagement. This will enable the team to confidently increase the weight of this pillar in future editions, as better data becomes available. Furthermore, the importance of balanced profiles highlights the inherent challenge of constructing a composite indicator in a nascent field like AI, where the compensatory properties of arithmetic aggregation remain vital for contextualizing early-stage development.

## C. Supporting Materials and Data Availability

To support transparency, reproducibility, and future research, the materials used in the research and scoring of the Global Index on Responsible AI are publicly available through the GIRAI Public Repository. The repository includes:

- **Global Survey Methodology** — provides a general overview of the survey structure and design, serving as the link between the questionnaires and the resulting dataset.
- **Researchers Handbook** — provides researchers with the conceptual framework, index structure, and question-by-question guidance for identifying, assessing, and documenting evidence for each indicator.

- **Data Collection Manual** — describes the data collection tool, roles and responsibilities, and the communication and quality-review process followed throughout data collection.
- **Indicator Narratives** — detailed guidance on the scope and interpretation of each indicator.
- **Application of the European Union (EU) Frameworks in the GIRAI** — a document detailing the application of EU frameworks within the GIRAI.
- **Secondary Data Sources Metadata** — metadata on the secondary data sources incorporated into the index, including the rationale for their inclusion, their source, the original series used and processing details.
- **Scripts for Secondary Data Processing** — scripts used to process secondary data.
- **GIRAI Scoring Script** — used to calculate indicator, dimension, and overall scores.
- **GIRAI Global Survey Dataset** — the primary data collected through the survey.
- **GIRAI Global Survey Data Dictionary** — variable-level documentation of the Global Survey Dataset, including variable names, definitions, values, and units.
- **GIRAI Scores and Rankings** — the GIRAI country-level scores and rankings at the indicator, dimension, pillar, and index level.